\documentclass[%
 reprint,
superscriptaddress,
longbibliography,
 amsmath,amssymb,
 aps,
floatfix,
]{revtex4-2}

\usepackage{graphicx}
\usepackage{dcolumn}
\usepackage{bm}
\usepackage{braket}
\usepackage{siunitx}

\usepackage{capt-of} %
\usepackage{hyperref}

\begin{document}

\title{Observation of anisotropy-independent magnetization dynamics in spatially disordered Heisenberg spin systems}

\author{T.~Franz}
\email{These authors contributed equally to this work. }
\affiliation{Physikalisches Institut, Universit\"at Heidelberg, Im Neuenheimer Feld 226, 69120 Heidelberg, Germany}
\affiliation{Max-Planck-Institut für Quantenoptik, Hans-Kopfermann-Str. 1, Garching, Germany}
\author{S.~Geier}
\email{These authors contributed equally to this work. }
\affiliation{Physikalisches Institut, Universit\"at Heidelberg, Im Neuenheimer Feld 226, 69120 Heidelberg, Germany}
\author{C.~Hainaut}
\affiliation{Physikalisches Institut, Universit\"at Heidelberg, Im Neuenheimer Feld 226, 69120 Heidelberg, Germany}
\affiliation{Universite de Lille, CNRS, UMR 8523-PhLAM-Laboratoire de Physique des Lasers Atomes et Molecules, F-59000 Lille, France}
\author{A.~Braemer}
\affiliation{Physikalisches Institut, Universit\"at Heidelberg, Im Neuenheimer Feld 226, 69120 Heidelberg, Germany}
\author{N.~Thaicharoen}
\affiliation{Physikalisches Institut, Universit\"at Heidelberg, Im Neuenheimer Feld 226, 69120 Heidelberg, Germany}
\affiliation{Department of Physics and Materials Science, Faculty of Science, Chiang Mai University, Chiang Mai 50200, Thailand}
\author{M.~Hornung}
\affiliation{Physikalisches Institut, Universit\"at Heidelberg, Im Neuenheimer Feld 226, 69120 Heidelberg, Germany}

\author{E.~Braun}
\affiliation{Physikalisches Institut, Universit\"at Heidelberg, Im Neuenheimer Feld 226, 69120 Heidelberg, Germany}

\author{M.~G\"{a}rttner}
\affiliation{Institut für Festkörpertheorie und Optik,  Friedrich-Schiller-Universität Jena, Max-Wien-Platz 1, 07743 Jena, Germany}
\author{G.~Z\"urn}  
\affiliation{Physikalisches Institut, Universit\"at Heidelberg, Im Neuenheimer Feld 226, 69120 Heidelberg, Germany}
\author{M.~Weidem\"uller}
\email{weidemueller@uni-heidelberg.de}
\affiliation{Physikalisches Institut, Universit\"at Heidelberg, Im Neuenheimer Feld 226, 69120 Heidelberg, Germany}

\date{\today}

\begin{abstract}
An important step towards a comprehensive understanding of far-from-equilibrium dynamics of quantum many-body systems is the identification of unifying features that are independent of microscopic details of the system. We experimentally observe such robust features in the magnetization relaxation dynamics of disordered Heisenberg XX-, XXZ- and Ising Hamiltonians. We realize these Heisenberg spin models with tunable anisotropy parameter and power-law interactions in an ensemble of Rydberg atoms by encoding the spin in suitable Rydberg state combinations. We consistently observe stretched-exponential relaxation of magnetization for all considered spin models, collapsing onto a single curve after appropriate rescaling of time. This robust short-time relaxation behavior is explained by a perturbative treatment that exploits the strong disorder in pairwise couplings, which leads to a description in terms of approximately independent pairs of spins.
In numerical simulations of small systems, we show that these pairs of spins constitute approximate local integrals of motion, which remain at least partially conserved on a timescale exceeding the duration of the relaxation dynamics of the magnetization.
\end{abstract}

\maketitle

\section{Introduction}

\begin{figure*}[t]
    \centering
    \includegraphics[width=0.8\textwidth]{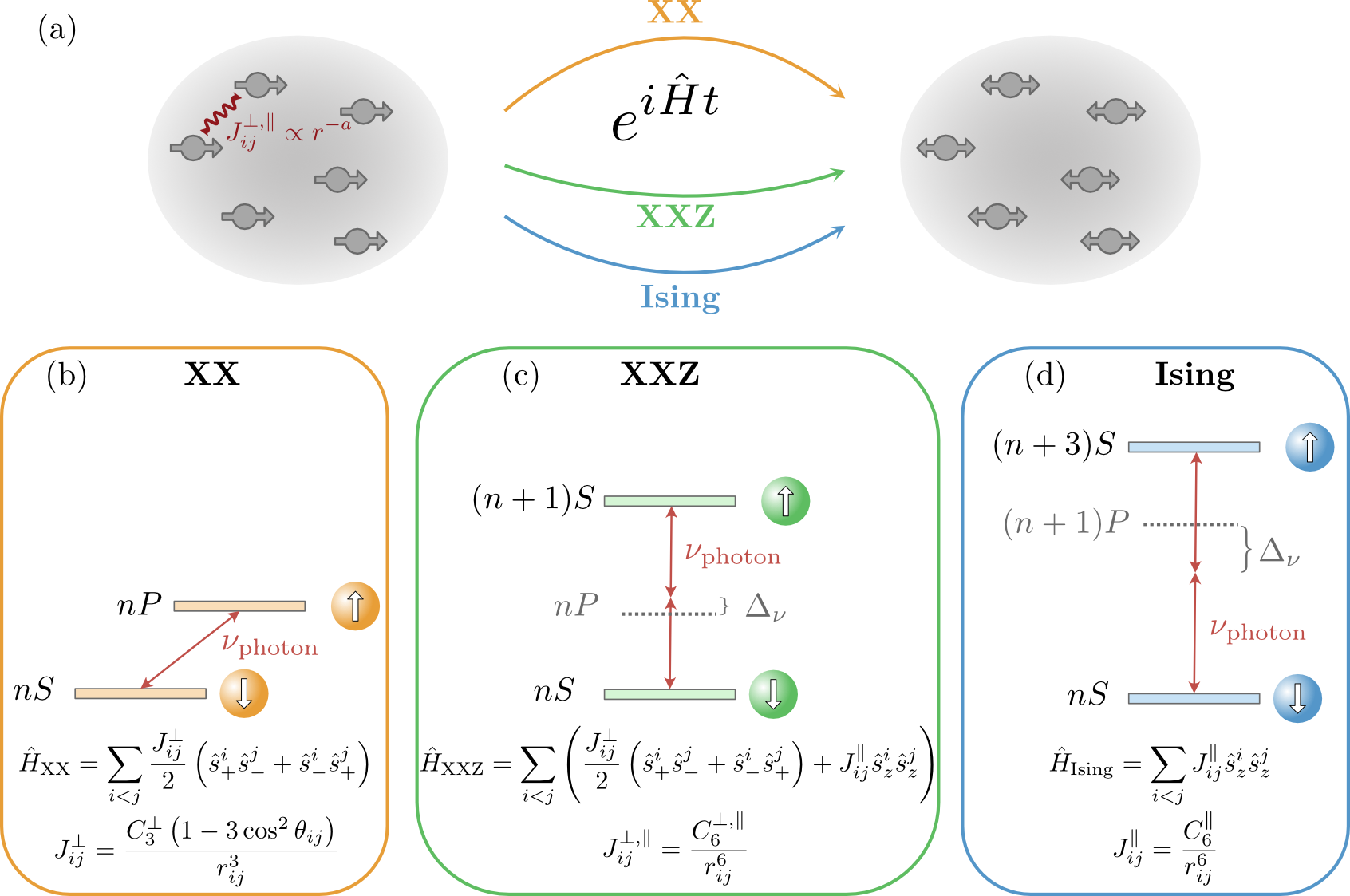}
    \caption{Rydberg quantum simulator platform. (a) Illustration of out-of-equilibrium disordered spin systems relaxing with respect to different Hamiltonians. (b) Illustration of the experimental realization of a Heisenberg XX Hamiltonian by coupling a Rydberg $\ket{nS}$ state to a $\ket{nP}$ state, possessing opposite parity. The interaction is of dipolar nature and falls off as $r_{ij}^{-3}$. Coupling two Rydberg states with the same parity results in a Heisenberg XXZ Hamiltonian for state combinations $\ket{nS}$ and $\ket{(n+1)S}$ (c), while state combinations $\ket{nS}$ and $\ket{(n+3)S}$ results in a Ising Hamiltonian (d). In the two latter cases, the interactions are of van der Waals nature with a $r_{ij}^{-6}$ dependence. 
    }
    \label{fig:figure1}
\end{figure*}

Far-from equilibrium dynamics of isolated quantum systems after a quench displays a wide range of emergent phenomena, such as dynamical phase transitions~\cite{Heyl2013DynamicalQuantum, Zhang2017ObservationManybody}, quantum many-body scars~\cite{Bernien2017ProbingManybody,serbyn_quantum_2021,bluvstein_quantum_2022} and many-body localization (MBL)~\cite{abanin_colloquium_2019,smith_many-body_2016,schreiber_observation_2015,lukin_probing_2019,choi_exploring_2016}. The time evolution of these systems generally depends strongly on the type of interactions and the distribution of interaction strengths between the particles~\cite{Polkovnikov2011ColloquiumNonequilibrium}. A notable exception are systems showing (metastable) prethermal phases, where relaxation dynamics can show universal behavior, i.e. the dynamics become independent of details of the microscopic model~\cite{ueda_quantum_2020,peng_floquet_2021,prufer_observation_2018,rubio-abadal_floquet_2020,eigen_universal_2018, Martin2022ControllingLocal}.

When considering the role of disorder for the dynamics of quantum many-body systems, a striking characteristic of the dynamics is that they can be non-ergodic~\cite{Benettin2014ErgodicityHow}, which is found for example in spin glasses where relaxation becomes extremely slow~\cite{Bernaschi2020StrongErgodicity} or in MBL systems where the dynamics might be completely frozen~\cite{Parameswaran2017EigenstatePhase}. Anomalously slow relaxation was also observed in disordered quantum spin systems that feature sub-exponential dynamics~\cite{Signoles2021GlassyDynamics,choi_depolarization_2017,kucsko_critical_2018,sommer_time-domain_2016,takei_direct_2016}. 
Remarkably, in all these different classical and quantum systems, in the strong disorder regime, the sub-exponential dynamics are well described by the same functional form, the stretched exponential law.
This raises the question of the origin of this robust behavior and whether it is affected by the modification of symmetry properties of the Hamiltonian. 

In classical systems, the answer to these questions is provided by the 
seminal work of Klafter and Shlesinger who found that a scale-invariant distribution of time-scales is the common underlying mathematical structure that induces stretched-exponential relaxation~\cite{klafter_relationship_1986}. Indeed, the authors proposed an intuitive understanding by considering the \textit{parallel channels} model where an ensemble of initially fully polarized spins are coupled to an external bath at a different strength sampled from a scale-invariant distribution. Due to the coupling to the bath, each spin decays exponentially on a different timescale. Thus, the global polarization of the system yields a stretched exponential form resulting from the averaging over all the spins.

For isolated quantum systems, where the dynamics are unitary, there is no notion of decay due to a bath. However, in a disordered system where the spins are randomly positioned in space, the interaction strengths between the spins can be distributed scale-invariantly. For example, it was shown analytically for the dynamics of the quantum Ising model that this scale-invariant distribution of coupling strengths induces a stretched exponential relaxation~\cite{Schultzen2022GlassyQuantum}. The derivation of the analytic solution is only possible because the Ising model features an extensive number of conserved quantities, i.e. it is integrable. 
For non-integrable models, where no analytic solution exists, generic mechanisms for describing the relaxation dynamics after a quantum quench remain largely unknown. Investigating the exact time evolution numerically is challenging due to the exponential growth of the Hilbert space with system size in quantum many-body systems. 
Semi-classical simulations, neglecting quantum effects beyond initial quantum fluctuations, suggest that non-integrable Heisenberg XYZ Hamiltonians present out-of-equilibrium dynamics that follows a stretched exponential law like the Ising model independent of their symmetry~\cite{Schultzen2022SemiclassicalSimulations}. 
An alternative route is implementing the desired unitary time evolution experimentally using quantum simulation experiments with tunable parameters, which is the approach we pursue here~\cite{Bloch2008ManybodyPhysics,Blatt2012, Georgescu2014QuantumSimulation, Gross2017QuantumSimulationsa}. 
 
In this work, we use different combinations of states of highly excited Rydberg atoms to realize different types of spin Hamiltonians thus making use of the full versatility of this platform~\cite{Browaeys2020ManybodyPhysics, Jaksch2000FastQuantum, Weimer2010RydbergQuantum, Barredo2015CoherentExcitation, Orioli2018RelaxationIsolateda, lukin_probing_2019, Signoles2021GlassyDynamics}. Rydberg atoms are ideally suited to study unitary quantum dynamics because the time scales of the interacting dynamics vastly exceed those of the typical decoherence mechanisms. 
We observe the relaxation dynamics of three different Heisenberg Hamiltonians: the integrable Ising model and the non-integrable XX and XXZ models with power-law interactions and positional disorder (see Fig.~\ref{fig:figure1}~(a)). For all models, we observe the same characteristic decay of magnetization, well-described by a stretched exponential function, which causes the data to collapse onto a single curve after the appropriate rescaling of time. 
We show that this robust behavior is directly linked to the presence of strong disorder which allows deriving an effective, integrable model consisting of pairs of spins. 

\section{Heisenberg Spin systems on a Rydberg-atom quantum simulator}

We consider an interacting spin-1/2 system described by the following Heisenberg Hamiltonian ($\hbar = 1$) 
\begin{equation}
    \hat{H} = \sum_{i<j}\left( J_{ij}^\perp/2 (\hat{s}_+^i\hat{s}_-^j +\hat{s}_-^i\hat{s}_+^j) + J_{ij}^\parallel \hat{s}_z^i\hat{s}_z^j  \right).
    \label{eq:SpinHamiltonian}
\end{equation}
Here, $\hat{s}_{\pm}^i = \hat{s}_x^i \pm i\hat{s}_y^i$, where $\hat{s}_\alpha^i (\alpha \in x,y,z)$ are the spin-1/2 operator of spin $i$ and $J_{ij}^{\perp, \parallel} = C_{a}^{\perp, \parallel}/r^{a}$. These types of Heisenberg XXZ Hamiltonians with disordered couplings feature a rich phenomenology of different phases and relaxation behaviors~\cite{Slagle2016DisorderedXYZ}. 
The Ising case, where $J_{ij}^{\perp} = 0$, features additional symmetries under local spin rotations $\hat{s}_z^{i}$ that commute with the Hamiltonian. 
which make the Ising model integrable. For $J_{ij}^{\perp} \neq 0$, $\hat{s}_z^{i}$ are no longer conserved and the Hamiltonian is non-integrable. 
We provide a comprehensive description of how to engineer this Hamiltonian with different combinations of Rydberg states in the appendix~\cite{Bijnen2013QuantumEngineering, Whitlock2017SimulatingQuantuma}. Fig. 1 illustrates the state combinations that can be used to realize the Heisenberg XX, XXZ, and Ising models. For the rest of this work, the three spin models are realized by state combinations $\ket{61S} - \ket{61P}$ (XX, $J^{\parallel}/J^{\perp} = 0$, $a=3$), $\ket{61S} - \ket{62S}$ (XXZ, $J^{\parallel}/J^{\perp} = -0.7$, $a=6$) and $\ket{61S} - \ket{64S}$ (Ising, $J^{\parallel}/J^{\perp} = -400$, $a=6$).

\section{Experimental observation of model independent relaxation dynamics}

The experiment starts with trapping Rubidium-87 atoms loaded in a crossed dipole trap at a temperature of \SI{20}{\micro\kelvin} (see appendix for experimental details). The atoms are excited from the ground $\ket{g} =  \ket{5S_{1/2}, F=2, m_F=2}$ to the Rydberg state $\ket{61S_{1/2}, m_j=0.5}$ by a two-photon transition with red (\SI{780}{\nano\meter}) and blue (\SI{480}{\nano\meter}) lasers that are detuned by $2\pi\cdot\SI{98}{\mega\hertz}$ from the intermediate state $\ket{e} = \ket{5P_{3/2}, F=3, m_F=3}$. For this state, the Rydberg lifetime of \SI{100}{\micro\second} exceeds the duration of the spin experiment of \SI{30}{\micro\second}. The excitation process leads to a three-dimensional cloud of N$\approx 80-250$ Rydberg atoms that are distributed randomly. The Van-der-Waals interaction during the excitation process imposes a minimal distance of $r_{\rm{bl}}\approx\SI{10}{\micro\meter}$ between the spins (Rydberg blockade effect). The state $\ket{61S_{1/2}, m_j=0.5}$ is the $\ket{\downarrow}$ state of all three different spin systems, the main difference is the second Rydberg states that is addressed by choosing proper microwave coupling using an AWG setup (see appendix for details). 

After having excited the ground state atoms to the down spin state, we implement a Ramsey protocol in our Rydberg experiment. To initialize the dynamics a first $\pi$/2-microwave pulse is performed, which sets the whole system is the state $\ket{\rightarrow}^{\otimes\rm{N}} = 1/\sqrt{2}(\ket{\uparrow}+\ket{\downarrow})^{\otimes\rm{N}}$ and we let the system evolve over \SI{30}{\micro\second}. A second $\pi$/2-pulse at a different readout phase followed by optical de-excitation and field ionization allows a tomographic measurement of the $x$ magnetization $\braket{\hat{S}_x} = \sum_i \braket{\hat{s}_x^i}$~\cite{Signoles2021GlassyDynamics}. 

\begin{figure*}[t]
    \centering
    \includegraphics[width=0.8\textwidth]{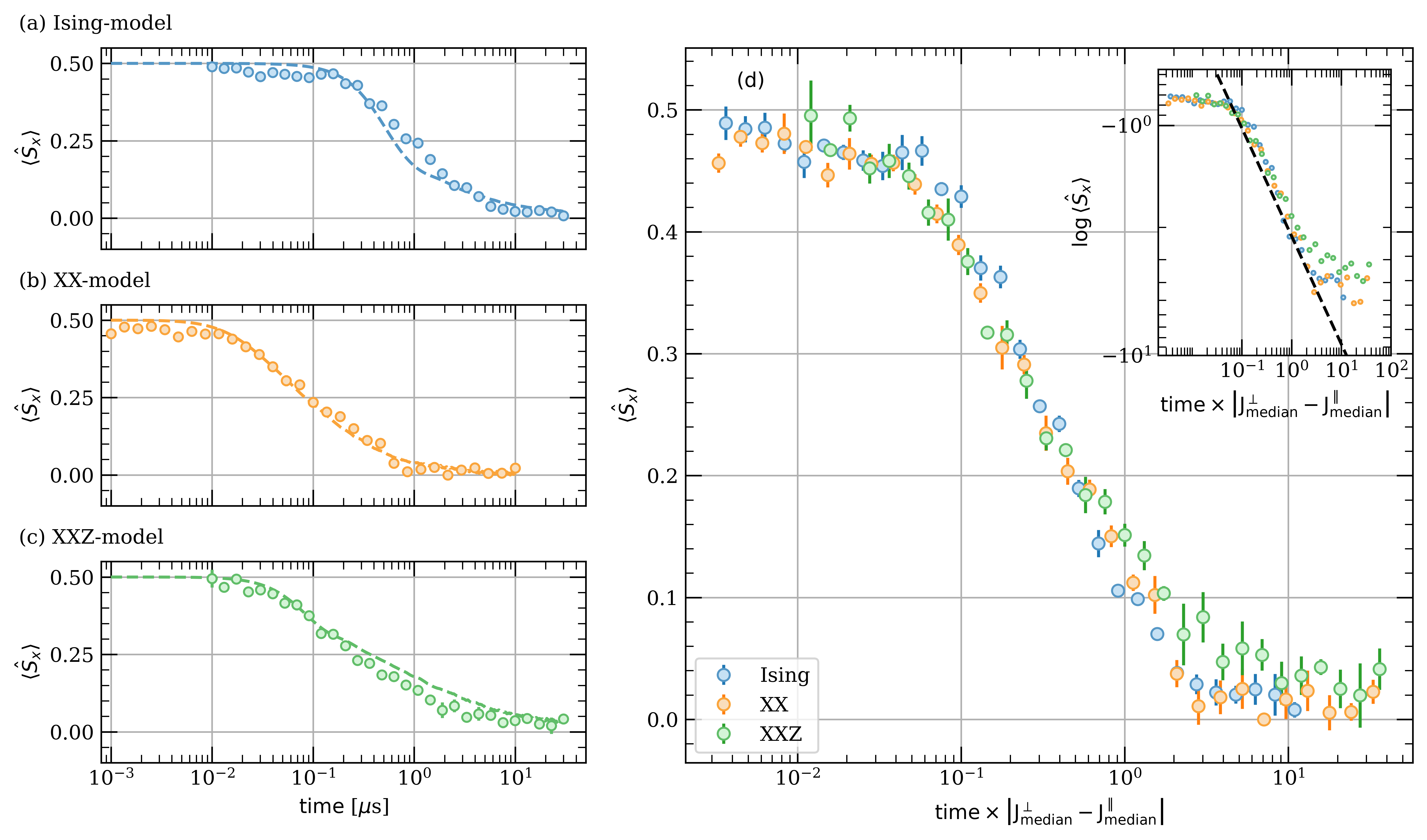}
    \caption{Relaxation dynamics of disordered quantum spin systems. Magnetization dynamics as a function of time for the Ising-model (a), the XX-model (b), and the  XXZ-model (c). The dashed lines stem from DTWA simulations. (d) Magnetization dynamics of the three models as a function of the time rescaled by the typical interaction strength $2\pi |J_{\rm{median}}^{\perp} - J_{\rm{median}}^{\parallel}|$ = \SI{2.3}{\mega\hertz} (Ising-model), \SI{21}{\mega\hertz} (XX-model), \SI{7.6}{\mega\hertz} (XXZ-model). Inset: Data points of (c) plotted on as loglog vs log. The dashed line is a guide to the eye, indicating a stretched exponential relaxation with $\beta=0.5$. The error bars denote the standard error of the mean.}
    \label{fig:figure2}
\end{figure*}

The resulting relaxation dynamics of the Ising, Heisenberg XX, and XXZ models are shown in Fig.~\ref{fig:figure2}~(a-c). At early times, the magnetization seems to be almost perfectly conserved at $\braket{\hat{S}_x} = 0.5$ before the relaxation begins. This effect is attributed to the Rydberg blockade that induces a maximal interaction strength that determines the system's fastest time scale. For each model, the system relaxes to zero magnetization, which can be understood by considering symmetry arguments: Indeed, the magnetization can be rewritten using the commutator relation for Pauli matrices $\langle \hat{S}_x\rangle = -i\langle [\hat{S}_y, \hat{S}_z]\rangle$. The latter term vanishes for each eigenstate $\ket{\phi}$ of the XXZ Hamiltonian because each eigenstate is also an eigenstate of $\hat{S}_z \ket{\phi} = \sum_i \hat{s}_z^{(i)} \ket{\phi} = S_z \ket{\phi}$ due to the global U(1) symmetry leading to $\langle [\hat{S}_y, \hat{S}_z]\rangle = S_z \braket{[\hat{S}_y, 1]} = 0$. The timescale of the dynamics occurring within less than \SI{10}{\micro\second} is comparable with the typical interaction strengths in the megahertz regime depending on the realized Heisenberg model (details on the distribution of interaction time scales can be found in the appendix).

To compare the relaxation curves to numerical predictions, the spatial distribution of Rydberg spin positions needs to be modeled realistically. We use a hard-sphere model where each Rydberg excitation is described by a superatom~\cite{Lukin2001DipoleBlockade} with a given blockade radius and effective Rabi frequency~\cite{Signoles2021GlassyDynamics}. For more details on the parameters of the models, see the appendix. 
We simulate the exact time-evolution of the experiment using the Discrete Truncated Wigner Approximation (DTWA)~\cite{Schachenmayer2015ManyBodyQuantum}. DTWA simulations agree well with the experimental data as shown in Fig.~\ref{fig:figure2}~(a-c). The small deviation between simulations and experiments can be mostly attributed to an inaccuracy of the atom distribution obtained from the simplified excitation model (see appendix).

The dynamics under the three different spin model in Fig.~\ref{fig:figure2}~(a-c) look strikingly similar in a log-linear plot. Indeed, by rescaling time with the characteristic timescale of each system given by $|J_{\rm{median}}^{\perp} - J_{\rm{median}}^{\parallel}|$, all relaxation curves coincide within the experimental errors. Here, 
\begin{equation}\label{eq:Jmedian}
    J_{\rm{median}}^{\perp, \parallel} = \mathrm{median}_j \max_i |J_{ij}^{\perp, \parallel}|
\end{equation}
is the median of the nearest neighbor interaction strengths. This choice of typical interaction time scale is motivated by the oscillation frequency of a single pair of interacting spins governed by \eqref{eq:SpinHamiltonian}, which will be further discussed in the following section. 
The striking collapse allows us to infer the functional form of the relaxation dynamics of the non-integrable models:
For the Ising model, it is known that the relaxation follows exactly the stretched exponential law $e^{-(t/\tau)^\beta}$~\cite{Schultzen2022GlassyQuantum} with stretching exponent $\beta$ and timescale $\tau$. The logarithm of the stretched exponential law is a power law. Plotted on a double logarithmic scale, this power-law becomes a linear function (dashed line in the inset of Fig.~\ref{fig:figure2}~(d)). In this representation, the rescaled experimental data also show a linear behavior. This confirms the hypothesis that the stretched exponential law is the unifying description of the magnetization relaxation for the integrable quantum Ising model as well as the non-integrable XX and XXZ Hamiltonians in the strongly disordered regime. We note that the dynamics are only robust with respect to a parameter of the microscopic model, the anisotropy $J^{\parallel}/J^{\perp}$, whereas the macroscopic geometry and also the dimension of the cloud may lead to different dynamics (see appendix). In addition, we also measured the relaxation dynamics for various initial states (for one Hamiltonian) possessing different magnetization and again find similar relaxation dynamics at late times (see appendix~\ref{app:initial_states}).

\section{Approximate description through strongly interacting pairs}
In order to understand the regime where we have observed robust relaxation dynamics, we aim for a simplified model that includes only the relevant time scales of the system. To identify these, we exploit the strongly disordered nature of the system by adopting a perturbative approach in the spirit of the strong disorder renormalization group (SDRG) where the strongest coupling is integrated out iteratively \cite{Vosk2013ManyBodyLocalization, Pekker2014HilbertGlassTransition, Vasseur2015QuantumCriticality, Vasseur2016ParticleholeSymmetry}. 

\begin{figure}[!htbp]
    \centering
    \includegraphics[width=\linewidth]{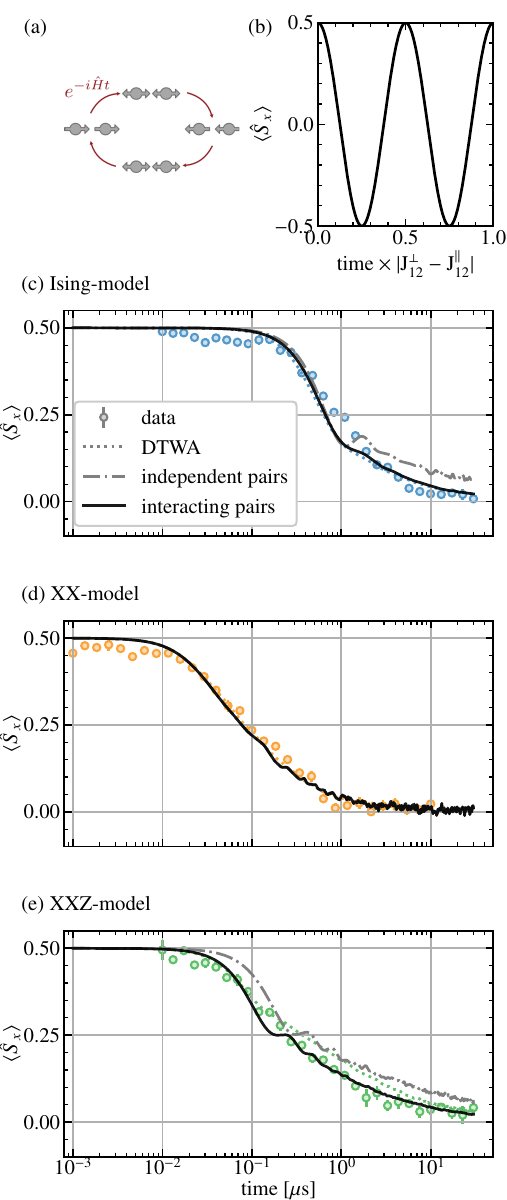}
    \caption{Effective description by localized pairs. (a) Illustration of the oscillation of a single pair under an arbitrary XXZ Hamiltonian. A fully polarized state     $\ket{\rightarrow\rightarrow}$ (left) evolves via the maximally entangled Bell state $1/\sqrt{2}\left( \ket{\rightarrow \rightarrow} + \ket{\leftarrow\leftarrow} \right)$ (top) to the state $\ket{\leftarrow\leftarrow}$ (right). Then, it returns to the origin via the other Bell state  $1/\sqrt{2}\left( \ket{\rightarrow \rightarrow} - \ket{\leftarrow\leftarrow} \right)$ (bottom).
    (b) Oscillation of the magnetization for a single pair initialized in $\ket{\rightarrow\rightarrow}$.
    (c-e) Comparison of the relaxation dynamics obtained by the pair approximation with/without effective Ising terms (solid black line/grey dash-dotted line) and with DTWA (dotted line) and the experimental data of Fig.~\ref{fig:figure2} for Ising (c), XX (d) and XXZ model (e).}
    \label{fig:figure3}
\end{figure}

In our model, the strongest coupled spins define a pair of spins. Crucially, the coupling within the pair will be much larger than all other couplings affecting the pair. This allows one to treat the coupling between this pair and the rest of the system perturbatively.
To zeroth order, this pair of spins just decouples from the system and evolves independently. 
This elimination step, where we remove the strongest coupling, can be repeated within the rest of the system. For our initial state, each individual pair undergoes coherent dynamics between the fully polarized state in plus and minus $x$-direction (see Fig.~\ref{fig:figure3}a)~\cite{Scholl2022MicrowaveEngineering}. The resulting oscillation of the magnetization (shown in Fig.~\ref{fig:figure3}~(b)) is independent of the specific XXZ Hamiltonian. Only the frequency, given by $J_{ij}^{\perp} - J_{ij}^{\parallel}$, differs depending on the Ising and exchange interaction strengths. 
This independence is at the origin of the observed model independence of relaxation dynamics. 

With this model in hand, we can compute the time evolution of the magnetization by a simple average of cosine oscillations as shown by the grey dash-dotted lines (pair,non-interacting) in Fig.~\ref{fig:figure3}c-e. The resulting relaxation dynamics show good agreement with the experimental data. However, especially for the Ising- and XXZ-model, this model underestimates the timescales of the dynamics. This is somewhat expected, considering that the pair couplings found by iterative elimination are, on average, smaller than the nearest neighbor couplings. 

Taking the perturbative treatment to next order, one finds an effective Ising-like coupling between pairs, as derived recently in the appendix of \cite{braemer_pair_2022}. The effective Hamiltonian governing the dynamics was found to be
\begin{align}
    \hat{H}_{eff} &\approx \sum_{\langle i,j\rangle}\left( J_{ij}^\perp/2 (\hat{s}_+^i\hat{s}_-^j +\hat{s}_-^i\hat{s}_+^j) + J_{ij}^\parallel \hat{s}_z^i\hat{s}_z^j  \right). \notag\\
    &\quad + \sum_{\langle i,j\rangle, \langle k,l\rangle} J^{eff}_{ijkl} \hat{s}_z^{(i)(j)}\hat{s}_z^{(k)(l)}
    \\
    J^{eff}_{ijkl} &= J_{ik}^{\parallel}+J_{il}^{\parallel}+J_{jk}^{\parallel}+J_{jl}^{\parallel}
\end{align}
where $\langle i,j\rangle$ denotes the summation over paired spins $i$ and $j$ and $2\hat{s}_z^{(i)(j)} = \hat{s}_z^{i}+\hat{s}_z^{j}$.

Fortunately, this model is integrable and allows for derivation of an analytical solution for the evolution of $\langle \hat{S}_x(t)\rangle$ (see Appendix~\ref{app:analytical_depolarization}), which reads
\begin{align}\label{eq:time_evo}
    \braket{\hat{S}_x^{\rm{pair}}}(t) &= \frac{1}{N}\sum_{\langle i,j\rangle}\! \cos(\frac{1}{2}(J_{ij}^{\perp} - J_{ij}^{\parallel}) t) \prod_{\langle k,l\rangle}\! \cos^2(\frac{1}{8}J^{eff}_{ijkl}t).
\end{align}
The first factor in each term originates from the pair dynamic to zeroth order, as described previously. The other factors are reminiscent of the Emch-Radin solution for the Ising model and stem from the effective Ising interaction among the pairs. This effective Ising model of pairs captures the overall demagnetization dynamics remarkably well for all observed times (see Fig.~\ref{fig:figure3}c-e), yielding very similar (and in the case of XXZ, even better) results compared to dTWA.

From the analytical form of the time evolution, Eq.~\ref{eq:time_evo}, we find that many different oscillation frequencies contribute to each spin's magnetization dynamics. Most of these frequencies are very small, however, and do not contribute to the early-time dynamics. Thus, a reasonable ansatz for rescaling to make the dynamics collapse is to consider only the fastest frequency for each spin. Due to the highly disordered nature of our system, this strongest coupling will essentially always correspond to the closest neighboring spin. This explains the rescaling found from the experimental data with $\mathrm{median}_i \max_j |J_{ij}^{\perp} - J_{ij}^{\parallel}|$.

\section{Separation of timescales in spin dynamics}

In the previous section, we revealed that the relaxation dynamics of a single-body observable is well captured by an ensemble of pairs with Ising-like interactions. This simple description in terms of pairs provides an integrable effective Hamiltonian which is valid not only at early times but agrees surprisingly well with the data over the entire relaxation process, which lasts for over three decades in time. In the following section, we will more quantitatively address the question of whether the magnetization of each pair is conserved by evaluating the pair autocorrelator given by $\langle \hat{S}_z^\text{pair}(t)\hat{S}_z^\text{pair} \rangle$, where $\hat{S}_z^\text{pair}=\hat{s}^i_z + \hat{s}^j_z$. If the pair picture is perfect or if the system is an Ising model, this quantity stays $\langle \hat{S}_z^\text{pair}(t)\hat{S}_z^\text{pair} \rangle = 1$. On the other hand, if the correlations in the system are fully decohered, the autocorrelator assumes its minimal value of $\langle \hat{S}_z^\text{pair}(t)\hat{S}_z^\text{pair} \rangle = \frac{2}{N}$ for a system of size $N$ due to symmetry constraints. 

\begin{figure}[!htbp]
    \centering
    \includegraphics[width=\linewidth]{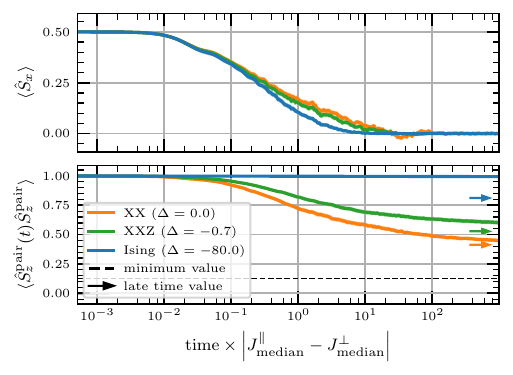}
    \caption{Dynamics of the pair magnetization autocorrelator. Numerical simulation of $N=16$ spins in $d=1$ with interaction exponent $a=2$ and a mean inter-spin distance $a_0=20r_b$. Line width shows statistical uncertainty from disorder averaging.}
    \label{fig:figure_autocorrelators}
\end{figure}
Our numerics presented in Fig.~\ref{fig:figure_autocorrelators} for $N=16$ spins in $d=1$ with interaction exponent $\alpha=2$ reveals three important points. Firstly, at $t\left| J_{\mathrm{median}}^{\parallel} - J_{\mathrm{median}}^{\perp} \right| \approx 0.2$ the global magnetization $\langle \hat{S}_x\rangle$ has decayed almost by half, while the pairs' magnetization autocorrelators are still close to 1. This justifies our simplistic pair picture and highlights the regime of universal dynamics. Secondly, at intermediate times up to $10^2$, the global magnetization relaxes fully to zero while the autocorrelator still features slow dynamics. This illustrates the existence of two timescales. Observing two distinct timescales shows that the system has not yet reached thermal equilibrium once the magnetization has relaxed to zero~\cite{choi_depolarization_2017,Martin2022ControllingLocal} but rather hints at prethermal behavior~\cite{Langen2016PrethermalizationUniversal,prufer_observation_2018,eigen_universal_2018,ueda_quantum_2020}. This generally means, that a system does not directly relax to its 'true' thermal state, but instead reaches a prethermal state. This is still a thermal state but with respect to a different, prethermal Hamiltonian which in our case only contains mostly Ising-like interactions among pairs. At very late times, this prethermal description ceases to be a reliable description, but even in the infinite time limit (derived by the diagonal ensemble and indicated by arrows in FIG. 4), the pair autocorrelator remains at  $\approx 1/2$ which is significantly above the lower limit of $\frac{2}{N} = 1/8$. This indicates that our integrable pair model is still a reasonable description of the system even at late times.

\section{Conclusion}

Our work demonstrates the ability of Rydberg atom quantum simulators to synthesize a variety of many-body Hamiltonians on a single experimental platform. By choosing the appropriate state combination, we realized XX-, XXZ-, and for the first time, a quantum Ising model within the Rydberg manifold. This versatility of the platform has enabled us to directly study and compare the relaxation dynamics of three different quantum spin systems far-from-equilibrium. 

The central finding of this study is the robustness of the functional form of the relaxation curves with respect to parameter changes in the microscopic spin model, even across models featuring different symmetries, and the choice of initial state (cf. Appendix~E). This discovery raises a fundamental question about the universality of relaxation dynamics in spatially disordered spin systems. To address this question comprehensively, we presented an approximate description of the system based on pairs of spins, exhibiting excellent agreement with both numerical simulations and experimental data. Moreover, this effective model is integrable and thus features an extensive number of conserved quantities allowing for an exact solution. 

To assess the quality of the effective model, we studied the decay of these effectively conserved quantities in small systems via exact methods. We found them to decay on a much slower timescale which might indicate that the system behaves prethermaly: On the early timescale, the effective pair model to lowest order holds, and thus the relaxation appears universal. 

The observed robustness hinges on a number of system properties: Firstly, universal relaxation is known to hold only in the strong disorder regime~\cite{Signoles2021GlassyDynamics}. Secondly, we expect the dynamics to depend on global parameters of the system like the spatial dimension $d$ and the range of interaction $\alpha$, which both determine the distribution of couplings $J_{i,j}^{\perp, \parallel}$ (e.g. the stretch power has been analytically derived to be $\beta = d/\alpha$ in the case of the Ising model~\cite{Schultzen2022GlassyQuantum}). Therefore, it is crucial to compare experimental data only where the distributions of interaction strengths are comparable such that the underlying universal behavior becomes evident (see also Appendix C where the distribution of coupling for the experiments shown in this article are shown). 

The accurate approximation of the relaxation dynamics by an integrable model of pairs indicates that the time evolution of disordered quantum spin systems cannot be viewed as direct thermalization.. Instead, even at later times when the global magnetization has completely relaxed to zero, the system can still exhibit local characteristics originating from quasi-conserved pairs of spins. 
In order to investigate the deviations from the pair model and, hence, from the prethermal state, future experiments will require single-site resolution~\cite{christakis_probing_2023}. 
Further investigations could also study the influence of the energy density of the initial state on the dynamics, indicative of a possible phase transition~\cite{Rademaker2020SlowNonthermalizing}.

\begin{acknowledgments}
We thank Andre Salzinger and Annika Tebben for important contributions to maintaining the experimental apparatus. Furthermore, we thank Hengyun Zhou, Nathan Leitao, and Leigh Martin for helpful discussions. This work is part of and supported by the Deutsche Forschungsgemeinschaft (DFG, German Research Foundation) under Germany’s Excellence Strategy EXC2181/1-390900948 (the Heidelberg STRUCTURES Excellence Cluster), within the Collaborative Research Centre “SFB 1225 (ISOQUANT),” the DFG Priority Program “GiRyd 1929,” the European Union H2020 projects FET Proactive project RySQ (Grant No. 640378), and FET flagship project PASQuanS (Grant No. 817482), and the Heidelberg Center for Quantum Dynamics. The authors acknowledge support by the state of Baden-Württemberg through bwHPC and the German Research Foundation (DFG) through grant no INST 40/575-1 FUGG (JUSTUS 2 cluster) and used the Julia programming language for most of the numerics~\cite{bezansonJuliaFreshApproach2017}. T. F. acknowledges funding by a graduate scholarship of the Heidelberg University (LGFG).
\end{acknowledgments}

\newpage

\appendix
\section{Engineering Heisenberg XXZ Hamiltonians by different combinations of Rydberg states}

In the following, we provide a comprehensive description of how to engineer this Hamiltonian with different combinations of Rydberg states [40, 41]. Especially, this gives us the opportunity to explain how to engineer an Ising Hamiltonian in a spin system realized by two different Rydberg states.

For general spin systems with global $U(1)$ symmetry, the coupling terms can be obtained by calculating the matrix elements of the interaction Hamiltonian $\hat{H}$. The Ising term
\begin{align}
    J_{ij}^{\parallel} = (E_{\uparrow_i\uparrow_j} + E_{\downarrow_i\downarrow_j}) - (E_{\downarrow_i\uparrow_j} + E_{\uparrow_i\downarrow_j})
    \label{eq:IsingTerm}
\end{align}
is defined as the energy difference between spins being aligned and being anti-aligned. Here, $E_{\alpha_i\beta_j} = \braket{\alpha_i\beta_j | \hat{H} | \alpha_i\beta_j}$ are the interaction energy of spin $i$ and $j$ with $\alpha, \beta \in [\uparrow, \downarrow]$. The exchange term is determined by 
\begin{align}
    J_{ij}^{\perp} = \braket{\downarrow_i\uparrow_j| \hat{H} | \uparrow_i\downarrow_j}.
    \label{eq:ExchangeTerm}
\end{align}
 
For a system consisting of states with opposite parity, such as $\ket{\downarrow} = \ket{nS}$ and $\ket{\uparrow} = \ket{nP}$ (see Fig.~\ref{fig:figure1}~(b)), where $n$ is the principal quantum number, the dominant coupling is a direct dipolar interaction which can be described by the Hamiltonian
\begin{align}
    \hat{H}_{\rm{DDI}} = \frac{\hat{\mathbf{d}}_i \cdot \hat{\mathbf{d}}_j - 3 \left(\hat{\mathbf{d}}_i \cdot \mathbf{e}_{r_{ij}}\right)\left(\hat{\mathbf{d}}_j \cdot \mathbf{e}_{r_{ij}}\right)}{r_{ij}^3}.
    \label{eq:H_ddi}
\end{align}
where $\hat{\mathbf{d}}_i$ is the dipole operator of atom $i$, $\mathbf{e}_{r_{ij}}$ is the unit vector connecting the two atoms and $r_{ij}$ their distance.
Mapped Eq.~\ref{eq:H_ddi} on the spin Hamiltonian of Eq.~\ref{eq:SpinHamiltonian}, the resulting interaction coefficient is
\begin{equation}
    J_{ij}^{\perp} = \frac{C_3^\perp (1-3 \rm{cos}^2 \theta_{ij})}{r_{ij}^3} .
\end{equation} Here, $\theta_{ij}$ is the angle between $\mathbf{e}_{r_{ij}}$ and the quantization axis and $C_3^\perp$ the coupling parameter~\cite{Barredo2015CoherentExcitation, Orioli2018RelaxationIsolateda}. The Ising term $J_{ij}^\parallel$ is zero since interaction energy shifts $E_{\alpha_i\beta_j}$ are dipole forbidden. Therefore, this is a way to realize an XX model as depicted in Fig. \ref{fig:figure1}~b). In this work, we have chosen $\ket{61S}$ and $\ket{61P}$ leading to $C_3^\perp/2\pi$= \SI{3.14}{\giga\hertz \micro\meter\cubed}.

In the case where the two chosen states possess the same parity, such as the two atoms being in the same state $\ket{nS}$, direct dipolar coupling is forbidden and the leading interaction is a second-order process through a virtually excited pair state $\ket{m}$ and can be described by 
\begin{align}
    \hat{H}_{vdW} = -\frac{1}{\hbar}\sum_m \frac{\hat{H}_{\rm{DDI}} \ket{m}\bra{m} \hat{H}_{\rm{DDI}}}{\Delta_{\nu}} \ .
    \label{eq:H_vdw}
\end{align}
Here, the Foerster defect $\Delta_\nu$ is the energy difference between the initial state and the virtually excited state $\ket{m}$. This Hamiltonian gives rise to power-law interactions $J_{ij} = C_6/r_{ij}^6$ that scales with $n^{11}$. Especially, this term is large if a pair state $\ket{m}$ with a small Foerster defect exists. 
Many experiments exploit these interactions to realize a spin system where the ground state is coupled to a single Rydberg state. These systems feature the Rydberg blockade effect and can be mapped on an Ising model~\cite{Jaksch2000FastQuantum, Weimer2010RydbergQuantum, Bernien2017ProbingManybody}. 

Similar interactions also exist for a spin system realized with two different Rydberg states $\ket{\downarrow }=\ket{nS}$ and $\ket{\uparrow }=\ket{(n+1)S}$. In this case, the Van-der-Waals Hamiltonian \eqref{eq:H_vdw} also induces a spin exchange term because the two Rydberg states are coupled via the intermediate pair state $\ket{m} = \ket{nP, nP}$ (see Fig.~\ref{fig:figure1}~(c)). 
In the case of $n=61$, both the Ising and exchange interactions terms are similar with $J^{\parallel}/J^{\perp} = -0.7$. Therefore, this spin system can be mapped onto an effective Heisenberg XXZ-Hamiltonian~\cite{Signoles2021GlassyDynamics}.
 
In order to realize an Ising Hamiltonian with two different Rydberg states, a state combination is needed where the exchange term \eqref{eq:ExchangeTerm} is small requiring a large Foerster defect $\Delta_{\nu}$ (see Fig.~\ref{fig:figure1}~(d)). This can be achieved by coupling $\ket{\downarrow }=\ket{nS}$ to $\ket{\uparrow }=\ket{(n+3)S}$. In this case, the largest contribution to the exchange term comes from $\ket{m} = \ket{(n + 1)P, (n+ 1)P}$. For example, for $n=61$, this spin system is characterized by a ratio of $J^{\parallel}/J^{\perp} = 400$, which is a good approximation to an Ising Hamiltonian ($J^{\perp} = 0$). 

\section{Experimental implementation of various spin models}

To realize the Heisenberg-XX model, a single-photon microwave transition at $2\pi\cdot\SI{16}{\giga\hertz}$ with a Rabi frequency of  $\Omega = 2\pi\cdot\SI{18}{\mega\hertz}$ couples the state $\ket{\downarrow}$ to $\ket{\uparrow} = \ket{61P_{3/2}, m_j=1/2}$. 
To implement the XXZ Hamiltonian, two microwave photons at $2\pi\cdot\SI{16}{\giga\hertz}$ couple to $\ket{\uparrow} = \ket{62S_{1/2}, m_j=1/2}$. Here, a single photon Rabi frequency of $\Omega = 2\pi\cdot\SI{48}{\mega\hertz}$ with a detuning $\Delta_{\nu} = 2\pi\cdot\SI{170}{\mega\hertz}$ leads to a two-photon Rabi frequency of $\Omega_{2\gamma} = 2\pi\cdot\SI{6.8}{\mega\hertz}$. 
To realize the Ising model, the state $\ket{61S}$ has to be coupled to $\ket{64S_{1/2}, m_j=0.5}$ but the detuning of $\Delta_{\nu} = 2\pi\cdot\SI{1.426}{\giga\hertz}$ is too large and prevents an efficient coupling of the states with two microwave photons of the same frequency $2\pi\cdot\SI{47}{\giga\hertz}$. Therefore, we combine two frequencies differing by $2\pi\cdot\SI{1.563}{\giga\hertz}$ such that the effective detuning to the intermediate state $\ket{62P}$ is $2\pi\cdot\SI{136}{\mega\hertz}$. For a single photon Rabi frequency of $\Omega = 2\pi\cdot\SI{30}{\mega\hertz}$ this results in an effective two-photon Rabi frequency of $\Omega_{2\gamma} = 2\pi\cdot\SI{3.3}{\mega\hertz}$ (see Fig.~\ref{fig:figure1}~(b-d) for the microwave photonic transitions).

\begin{center}
\begin{tabular}{||c| c c| c c||} 
 \hline
  Model & blue exc. $\sigma_{x,y}$ & red exc. $\sigma_{x,y}$ & GS $\sigma_x$ & GS $\sigma_{y,z}$  \\ [0.5ex] 
 \hline\hline
 Ising & 55$\mu$m & 1.5mm & 64$\mu$m & 45$\mu$m \\ 
 \hline
 XXZ & 55$\mu$m & 1.5mm & 64$\mu$m & 45$\mu$m \\
 \hline
 XX & 55$\mu$m & 1.5mm & 62$\mu$m & 47$\mu$m \\ [1ex] 
 \hline
\end{tabular}
\captionof{table}{Waists of the blue (480 nm) and red (780 nm) Rydberg excitation lasers used to realize the different models and the respective ground state cloud waists.}
\end{center}

\section{Distribution of interaction time scales in the spin system}

In the main text, we have highlighted that the typical timescale of the relaxation is given by the pair oscillation frequency $|J_{\parallel} - J_{\perp}|$. For the Heisenberg XXZ Hamiltonian, both exchange and Ising interactions exist. Therefore, another possibility of rescaling would only involve $J_{\perp}$ which would disregard the anisotropy $\delta = J_{\parallel}/J{\perp}$. In Fig.~\ref{fig:comparison_scaling}, we have compared both possibilities of rescaling time. The rescaling by the oscillation frequency shows a more precise collapse of the experimental data. This demonstrates that this frequency indeed determines the relevant timescale of the system. In addition, this indicates that the Rydberg interactions can be mapped onto the Heisenberg XXZ Hamiltonian with $\delta=-0.7$.

In Fig.~\ref{fig:simulations_density_blockade}, we show the sensitivity of the DTWA simulations to different densities and blockade radii. For most simulations, these parameters have only a small, quantitative effect on the simulated dynamics. A notable exception is the Ising system. Here, the Rydberg cloud is largely saturated and the blockade radius is the relevant length scale of the system. Therefore, a variation of the blockade radius changes drastically the early time dynamics. In contrast, the density of the sample featuring XX-interaction is low, therefore the blockade effect can be neglected. For the Heisenberg XXZ Hamiltonian, the simulations show that the blockade radius of \SI{8.3}{\micro\meter} fits the observed dynamics slightly better than the value of \SI{10}{\micro\meter} expected from the simplified excitation model assuming no phase noise of the laser. 

\begin{figure}[t]
    \centering
    \includegraphics[width=0.5\textwidth]{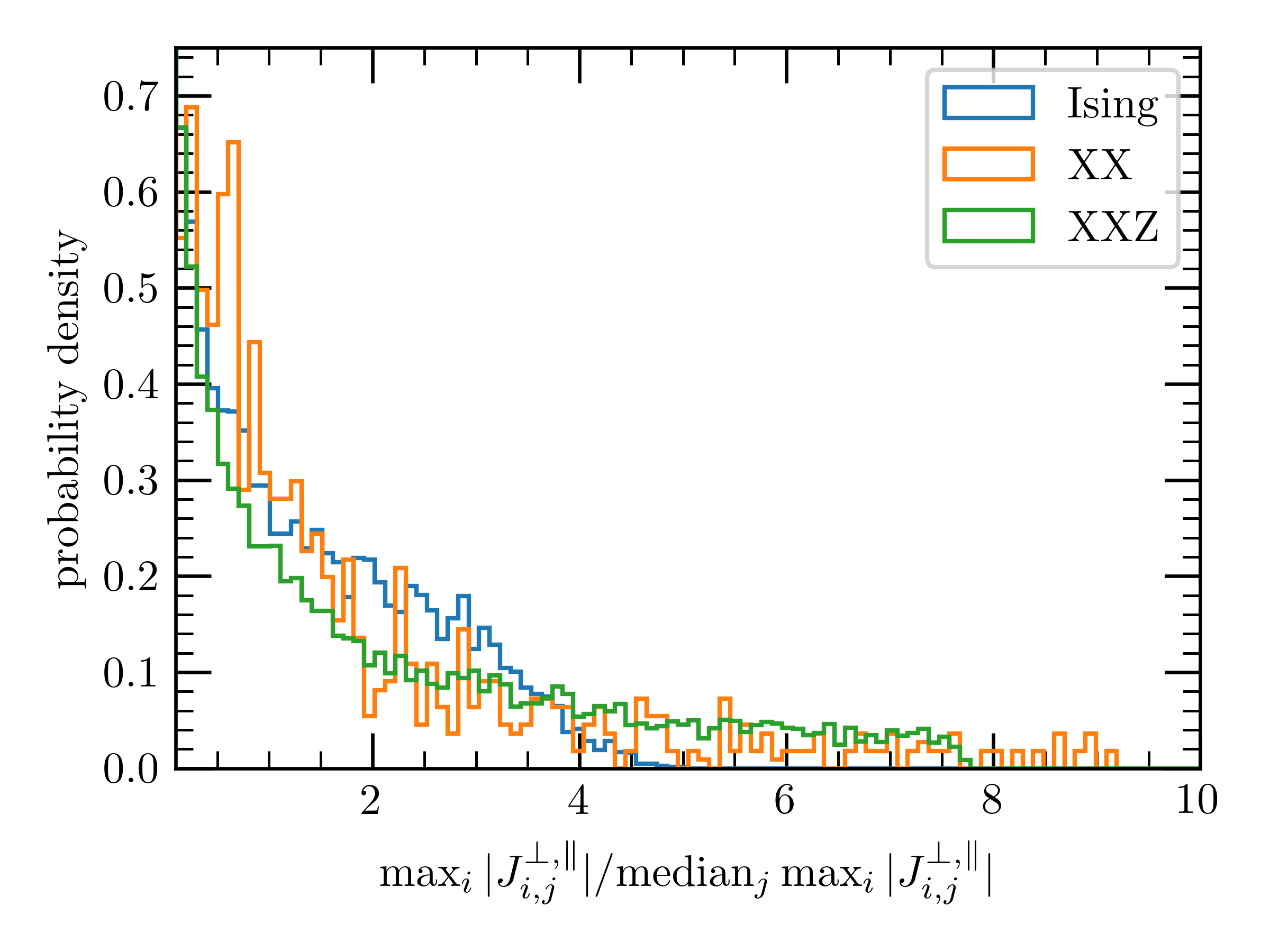}
    \caption{Histograms of the distribution of interaction strengths for the experimental data shown in the main text. The histograms are obtained by averaging over 50 distributions of interaction strengths, with a bin size of $0.1 \mathrm{median}_j \max_i|J_{i, j}^{\perp, \parallel}|$.
    }
    \label{fig:interaction_distribution}
\end{figure}

\begin{figure*}[t]
    \centering
    \includegraphics[width=0.8\textwidth]{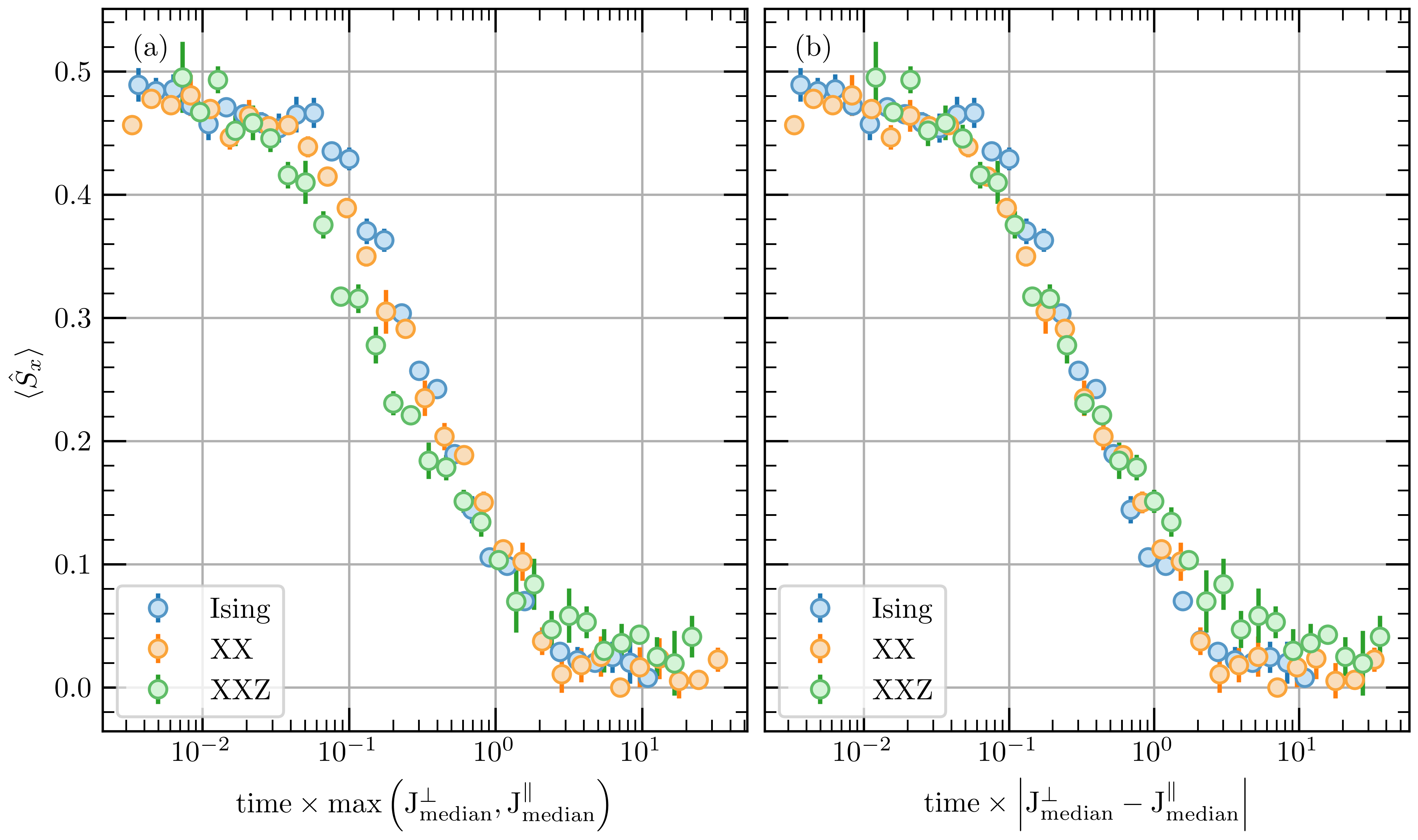}
    \caption{Comparison of the scaling behavior for rescaling time either by the median interaction matrix (a) or by the median of the pair oscillation frequency (b). $\max(J_{\rm{median}}^{\perp}, J_{\rm{median}}^{\parallel})$ is defined as $J_{\mathrm{median}}^{\perp}$ for the Heisenberg XX and XXZ model, and as $J_{\rm{median}}^{\perp}$ for the Ising model.}
    \label{fig:comparison_scaling}
\end{figure*}

\begin{figure*}[!htbp]
    \centering
    \includegraphics[width=0.8\textwidth]{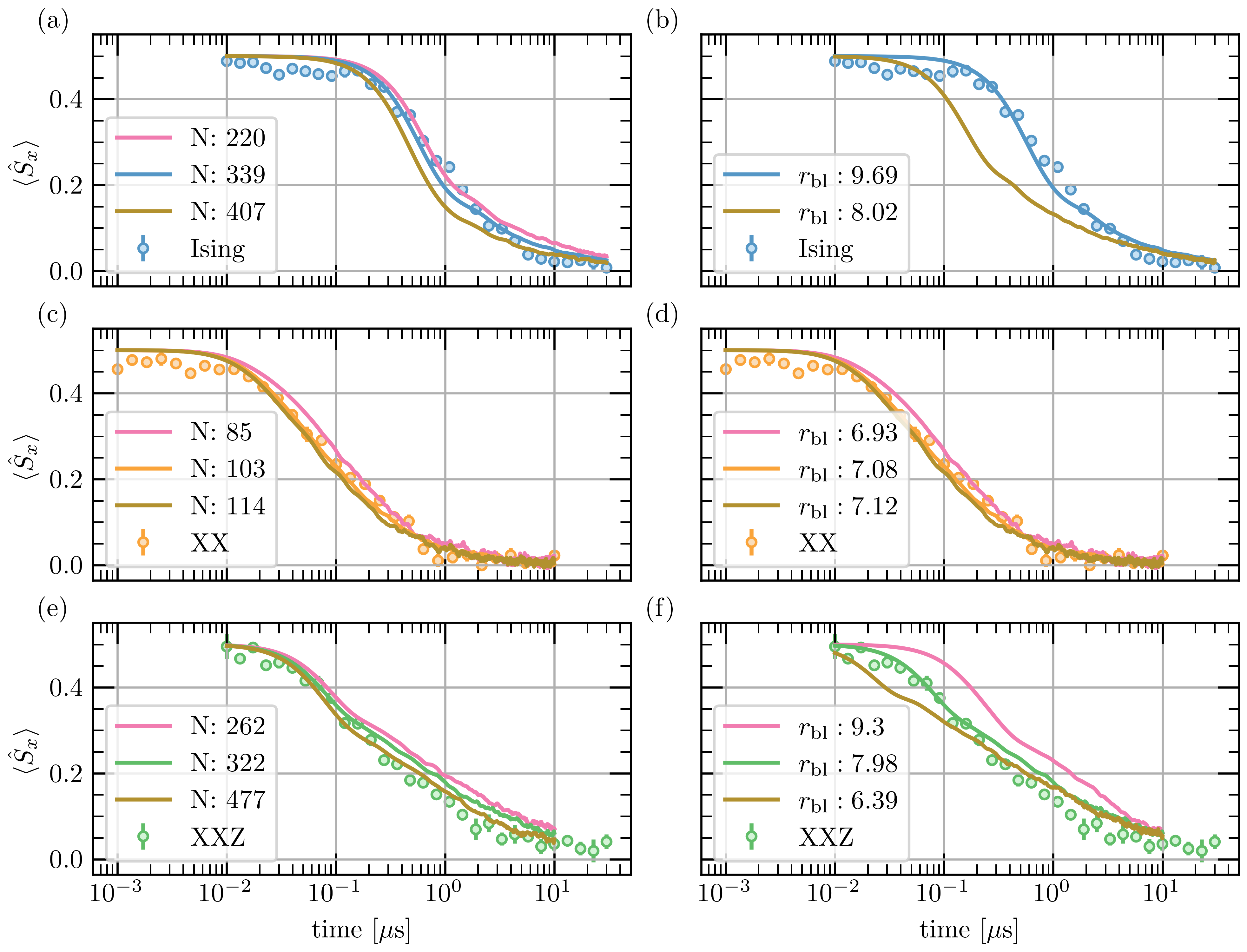}
    \caption{Influence of the density and the blockade radius on the DTWA simulations. Left column: Simulations for the same blockade radius as in the main text for different particle numbers $N$. Right column: Simulations for the same particle number and various blockade radii. In all simulations the geometry of the cloud is the same. }
    \label{fig:simulations_density_blockade}
\end{figure*}

Histograms showing the resulting distribution functions of couplings are shown in Figure~\ref{fig:interaction_distribution}.

\section{Influence of the Gaussian trap geometry on the relaxation dynamics}

The functional form of the relaxation dynamics in a strongly disordered spin system has been demonstrated to be independent of both the Rydberg blockade radius and the anisotropy of the Heisenberg XXZ Hamiltonian. However, the timescale of these dynamics is contingent upon the density and coupling constant. Consequently, there arises the necessity to rescale time by the median interaction strength $\mathrm{median}i\max_j J_{ij}$.

\begin{figure}[!htbp]
    \centering
    \includegraphics[width=\linewidth]{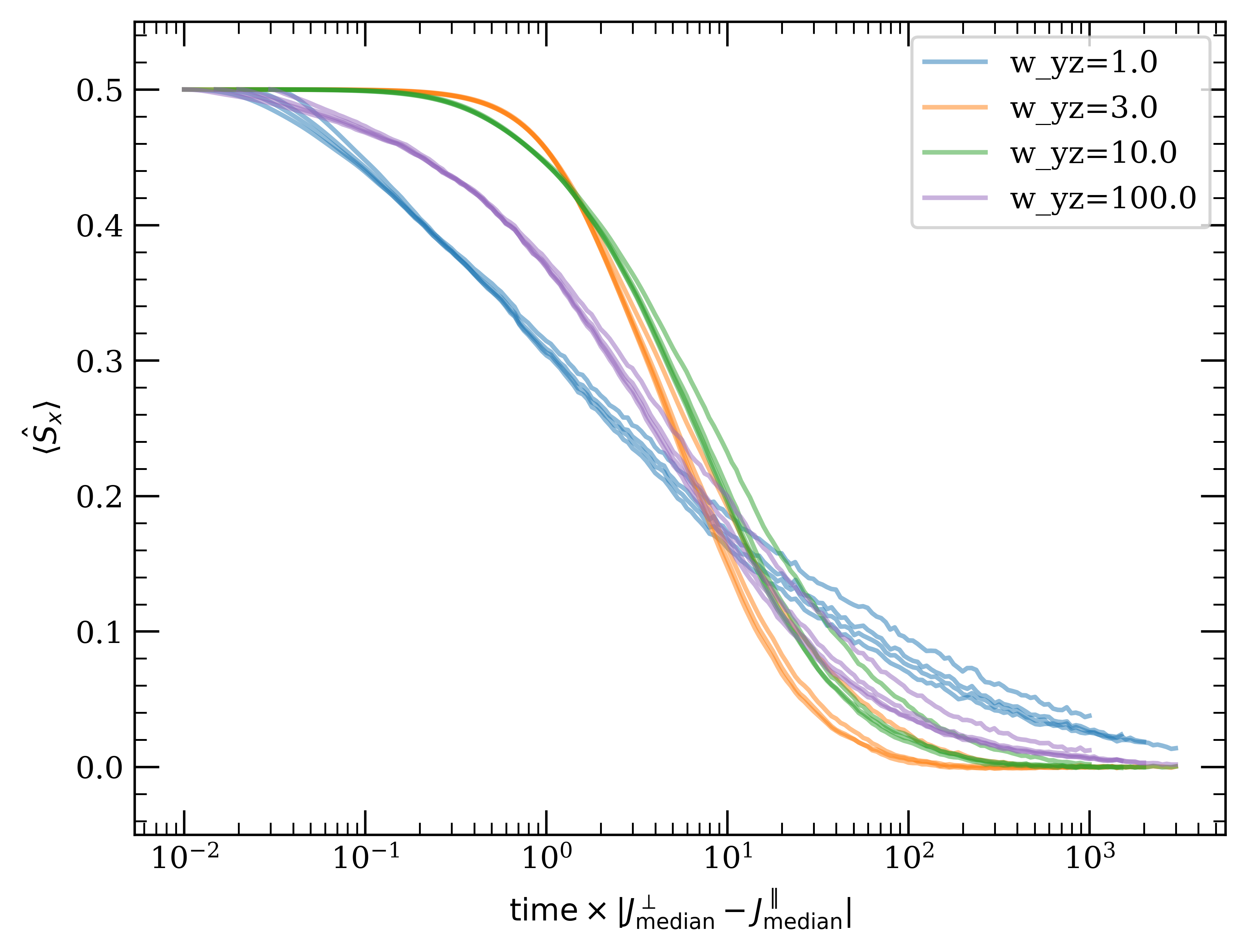}
    \caption{MACE simulations of the relaxation of the magnetization for four different geometries of the Gaussian cloud where the aspect ratio of the waist $\mathrm{w}_{x}$ in $x$ direction with respect to the waist $\mathrm{w}_{yz}$ in $y$ and $z$ direction is tuned. The product $\mathrm{w}_{x} \times \mathrm{w}_{yz}^2$ is fixed for all 4 geometries. For each geometry, we simulate the time evolution for different anisotropies $\frac{J^\parallel}{J^\perp} \in \{-2, -0.5, 0, 0.5, 2\}$.} 
    \label{fig:comparison_geometries}
\end{figure}

In the context of a Gaussian trap geometry, we conduct an averaging procedure over varying local densities $\rho$. Assuming local density approximation, we average over different local relaxation dynamics, each characterized by a stretched exponential function, sharing a common stretching exponent $\beta$, while exhibiting distinct timescales $\tau(\rho)$. The collective summation of these relaxation curves again manifests as a stretched exponential decay. However, the details of the stretching exponent $\beta$ depend on the shape of the Gaussian cloud (cf. Fig.~\ref{fig:comparison_geometries}).

Furthermore, finite-size effects come into play, with one-dimensional physics becoming relevant in an elongated cigar-shaped geometry and two-dimensional physics in a flat pancake geometry. Consequently, the measured stretched exponential does not align with the expected value of $\beta=d/\alpha$ (where $d$ represents the dimension and $\alpha$ signifies the range of interactions) as anticipated from semiclassical simulations~\cite{Schultzen2022SemiclassicalSimulations}. Instead, the observed value interpolates between various dimensions and exhibits slight variations in different experimental realizations when the trap geometry is altered.

Nevertheless, through a comparative analysis of experiments conducted in similar geometries, it remains feasible to investigate whether the dynamics are contingent upon the size of the blockade radius~\cite{Signoles2021GlassyDynamics} or the anisotropy parameter $\Delta$ of the Heisenberg XXZ Hamiltonian (as explored in this study).

\section{Relaxation under initial states with different magnetization}\label{app:initial_states}

\begin{figure}[!htbp]
    \centering
    \includegraphics[width=0.4\textwidth]{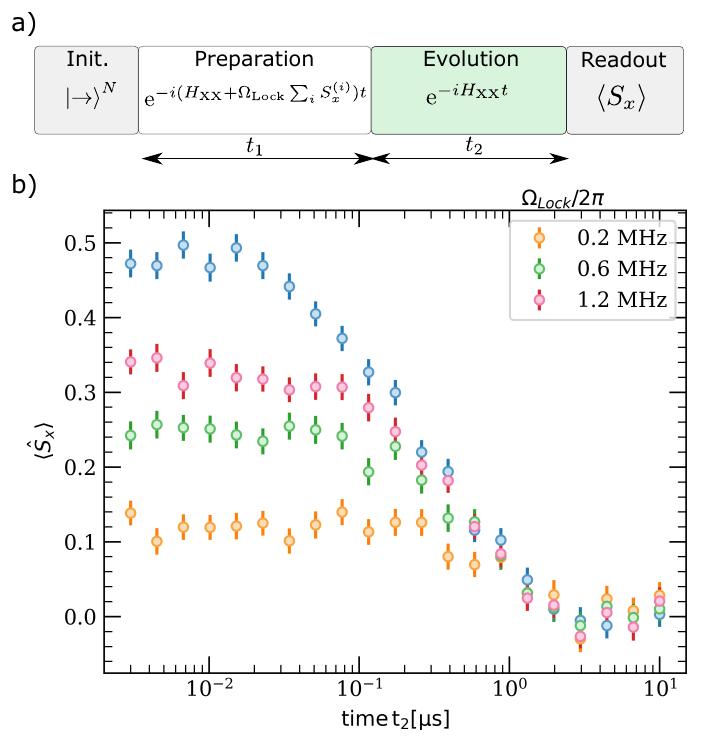}
    \caption{Relaxation dynamics of initial states with different magnetization. a) Experimental sequence consisting of an evolution under a spin locking field (preparation), followed by a measurement of the magnetization for the resulting relaxation dynamics for $t_2$ (evolution). b) Magnetization dynamics after different $\Omega_{\rm{Lock}}/2\pi$ applied in phase one.}
    \label{fig:locking}
\end{figure}

In order to test if the universal relaxation behavior, originating from a pair picture approximation is even consistent for the relaxation of more general states than the fully x-polarized state, we probe the relaxation for initial states with different magnetization.
The experimental protocol is shown in Fig.~\ref{fig:locking}a and is similar to the one used in \cite{Franz2022AbsenceThermalization}. It consists of the following steps. In the preparation, spins are initially polarized along the x-axis. A locking field $\Omega_{\rm{Lock}}$ which is also aligned along the x-axis is applied for a time $t_1$. During this time, as reported in \cite{Franz2022AbsenceThermalization}, the magnetization will relax and approximately settle to a constant non-zero value that depends on the strength of the locking field. In the evolution, we then turn off the locking field and measure the resulting relaxation of the x-magnetization.
The resulting relaxation over period $t_2$ is shown in Fig.~\ref{fig:locking}b. The magnetization starts with different values depending on the field strength applied during the preparation. We note that the locking time $t_2 = 2 \mu s$ is larger than the time it takes to directly relax to zero magnetization without phase one (blue points). We observed that for decreasing initial magnetization, the onset of the relaxation dynamics gets shifted to a later time (red, green, and yellow points). However, independent of this behavior, for later times, all curves overlap with the direct relaxation curve without field (blue points).

The observed dynamics can be understood within the pair approximation in the following way. During preparation, the locking field is only able to lock pairs with interactions smaller than the field strength $\Omega_{\rm{Lock}}$.  These pairs stay polarized while pairs with stronger interactions oscillate and dephase. As reported in \cite{Franz2022AbsenceThermalization}, magnetization takes an almost constant value. During the evolution, when the field is turned off, the relaxation timescale is given by the remaining pairs that were locked and now start to oscillate. This timescale is longer for small fields where only weakly interacting pairs remained locked during the preparation. The overlapping at a later time is due to the fact that these pairs are also locked under larger fields.
The data was take for $\ket{48S_{1/2},m_j=0.5}$ and $\ket{48P_{3/2},m_j=0.5}$.

\section{Derivation of depolarization dynamics}\label{app:analytical_depolarization}
The goal is to compute the expectation value of $\braket{\hat{S}_x(t)} = \frac{1}{N}\sum_i \braket{\hat{s}_x^{i}}$ starting from the $x$-polarized state $\ket{\psi_0}=\ket{\rightarrow}^{\otimes N}$ governed by the effective Hamiltonian derived in \cite{braemer_pair_2022}
\begin{align}
    \hat{H}_{eff} &= \sum_{\langle i,j\rangle}\left( J_{ij}^\perp (\hat{s}_x^i\hat{s}_x^j +\hat{s}_y^i\hat{s}_y^j) + J_{ij}^\parallel \hat{s}_z^i\hat{s}_z^j  \right). \notag\\
    &\quad + \sum_{\langle i,j\rangle, \langle k,l\rangle} J_{eff}^{ijkl} \hat{s}_z^{(i)(j)}\hat{s}_z^{(k)(l)}
\end{align}
where $\langle i,j\rangle$ denotes the summation over paired spins $i$ and $j$ and $2\hat{s}_z^{(i)(j)} = \hat{s}_z^{(i)}+\hat{s}_z^{(j)}$.

Without loss of generality, we assume that spins 1 and 2 form a pair and compute $\braket{\hat{s}_x^{1}(t)}$. The evolution of $\braket{\hat{S}_x(t)}$ then follows simply by linearity. First we notice that all the terms in $\hat{H}_{eff}$ commute with each other, allowing for direct computation of $\braket{\hat{s}_x^{1}(t)}$ by commuting $\hat{s}_x^{1}$ through the time evolution operators:
\begin{widetext}
    \begin{align}
        \hat{s}_x^1(t) &= e^{it\hat{H}_{eff}} \hat{s}_x^1 e^{-it\hat{H}_{eff}}\\
        &= e^{itJ^{\perp}_{12}(\hat{s}_x^1\hat{s}_x^2 +\hat{s}_y^1\hat{s}_y^2)} e^{itJ^{\parallel}_{12}\hat{s}_z^1\hat{s}_z^2} e^{it\hat{s}_z^{(1)(2)} \sum_{\langle k,l\rangle} J_{eff}^{12kl} \hat{s}_z^{(k)(l)}} \hat{s}_x^1 e^{-it\hat{s}_z^{(1)(2)} \sum_{\langle k,l\rangle} J_{eff}^{12kl} \hat{s}_z^{(k)(l)}} e^{itJ^{\parallel}_{12}\hat{s}_z^1\hat{s}_z^2} e^{-itJ^{\perp}_{12}(\hat{s}_x^1\hat{s}_x^2 +\hat{s}_y^1\hat{s}_y^2)}\\
        &= e^{itJ^{\perp}_{12}(\hat{s}_x^1\hat{s}_x^2 +\hat{s}_y^1\hat{s}_y^2)} e^{it\hat{s}_z^{1} \sum_{\langle k,l\rangle} J_{eff}^{12kl} \hat{s}_z^{(k)(l)}} e^{2itJ^{\parallel}_{12}\hat{s}_z^1\hat{s}_z^2} e^{-itJ^{\perp}_{12}(\hat{s}_x^1\hat{s}_x^2 -\hat{s}_y^1\hat{s}_y^2)} \hat{s}_x^1
    \end{align}
\end{widetext}
Now we can just expand the exponentials using the usual formula for the exponential of Pauli matrices (note that $\hat{s}_x^1\hat{s}_x^2 +\hat{s}_y^1\hat{s}_y^2$ is akin to $\hat{s}_x$ in a specific subspace) and take the expectation value with respect to the initial state to get the desired result:
\begin{align}
    \braket{\hat{s}_x^1(t)} = \frac{1}{2}\cos\left(\frac{J^{\perp}_{12} - J^{\parallel}_{12}}{2}t\right)\prod_{\langle k,l\rangle} \cos^2\left(\frac{J_{eff}^{12kl}}{8}t\right)
\end{align}

\clearpage
\bibliography{bibliography}

\end{document}